\documentclass[12pt]{article}
\usepackage{amsfonts,amssymb}

\def\on#1#2{\mathop{\vbox{\ialign{##\crcr\noalign{\kern2pt}
$\scriptstyle{#2}$\crcr\noalign{\kern2pt\nointerlineskip}
\kern-2pt$\hfil\displaystyle{#1}\hfil$\crcr}}}\limits}

\def\nn{ \nonumber }
\def\bq{ \begin{equation} }
\def\eq{ \end{equation} }
\def\ben{ \begin{eqnarray} }
\def\en{ \end{eqnarray} }
\def\frac#1#2{{#1\over #2}}
\def\dfrac#1#2{{\displaystyle{#1\over#2}}}

\begin{document}
\title{Integrable systems on
$so(4)$ related with $XXX$ spin chains with boundaries.}
\author{
A.V. Tsiganov, O.V. Goremykin\\
\\
\it\small  St.Petersburg State University, St.Petersburg, Russia}

 \date{}
\maketitle

\begin{abstract} {\small
We consider two-site $XXX$ Heisenberg magnets with  different
boundary conditions, which  are integrable systems on $so(4)$
possessing additional cubic and quartic integrals of motion. The
separated variables for  these models are constructed using the
Sklyanin method.

 }
\end{abstract}
\vskip0.8cm \noindent{ PACS numbers: 02.30.Ik, 02.30.Uu, 02.30.Zz,
02.40.Yy, 45.30.+s } \vglue1cm \textbf{Corresponding Author}: A.V.
Tsiganov, Department of Mathematical and Computational Physics,
Institute of Physics, St.Petersburg University, 198 904,
St.Petersburg, Russia,\\ e-mail: tsiganov@mph.phys.spbu.ru
 \newpage

\section{Introduction}

In classical mechanics a lot of integrable systems on $so(4)$ and
$e(3)$ algebras are related with the integrable spin chains. For
instance \cite{mag}, different Gaudin magnets coincide with the
Euler, Lagrange, Neumann and Clebsh systems on $e(3)$ and with the
Manakov, Steklov systems on $so(4)$ . The separated variables for
all these models may be derived from the separation of variables
for the $XYZ$ Gaudin model \cite{skl98}.

Some degenerate cases of the Heisenberg spin chain are connected
with the Goryachev-Chaplygin top \cite{skl85}, auxiliary symmetric
Neumann system and Kowalevski-Goryachev-Chap\-ly\-gin top
\cite{kuzts89}. The separated variables for all these models may
be derived from the separation of variables for the $XYZ$ spin
chain \cite{skl95}.

In this note we consider $XXX$ Heisenberg magnet with boundaries
\cite{skl88,soktsig2} using the Lax matrix for the standard
two-site $XXX$ Heisenberg magnet
\bq \label{Lax-XXX}
T(\lambda)=\left(\begin{array}{cc} \lambda-s_3+ i\delta_1& s_1 + is_2\\
s_1 - is_2& \lambda+s_3 + i\delta_1\end{array}\right)
\left(\begin{array}{cc} \lambda-t_3+ i\delta_2& t_1 + it_2\\
t_1 - it_2& \lambda+t_3 + i\delta_2\end{array}\right).
\eq
Here $\delta_i$ are numerical shifts of the spectral parameter
$\lambda$. Dynamical variables  $s_i,t_i$ are coordinates on
$so(4)=so(3)\oplus so(3)$ with the following Lie-Poisson brackets
\begin{equation} \label{2o3}
\bigl\{ s_i\,,s_j\,\bigr\}= \varepsilon_{ijk}\,s_k\,, \qquad
\bigl\{ s_i\,,t_j\,\bigr\}= 0\,,\qquad \bigl\{
t_i\,,t_j\,\bigr\}=\varepsilon_{ijk} t_k\,,
\end{equation}
where $\varepsilon_{ijk}$ is the totally skew-symmetric tensor.

Matrix $T(\lambda)$ (\ref{Lax-XXX}) defines representation of the
Sklyanin algebra
\bq
\{\,\on{T}{1}(\lambda),\,\on{T}{2}(\nu)\}= [r(\lambda-\nu),\,
\on{T}{1}(\lambda)\on{T}{2}(\nu)\,]\,, \label{rrpoi}
\eq
on generic symplectic leaves  of  $so(4)$. Here we use the
standard notations\, $\on{T}{1}(\lambda)= T(\lambda)\otimes
I\,,~\on{T}{2}(\nu)=I\otimes T(\nu)$ and $r$-matrix has the form
\bq
r(\lambda-\nu)=\dfrac{i}{\lambda-\nu}\,\left(\begin{array}{cccc}
 1 & 0 & 0 & 0 \\
 0 & 0 & 1 & 0 \\
 0 & 1 & 0 & 0 \\
 0 & 0 & 0 & 1
\end{array}\right)\,.\label{rr}
\eq
Applying the standard machinery \cite{skl88,soktsig2} to
$T(\lambda)$ (\ref{Lax-XXX}) one gets another integrable systems
with cubic and quartic additional integrals of motion. As a matter
of fact, the construction leads automatically to the separated
variables.

\section{Cubic integrals of motion}
The main property of the Sklyanin algebra (\ref{rrpoi}) is that
for {\it any} numerical matrix ${\cal K}$ coefficients of the
trace of matrix ${\cal K}T(\lambda)$ give rise the commutative
subalgebra
\[\{\mbox{\rm tr}\,{\cal K}T(\lambda)\,,\mbox{\rm tr}\,{\cal K}T(\nu)\}=0\,.\]
All the generators of this subalgebra are linear polynomials on
coefficients of entries $T_{ij}(\lambda)$, which are interpreted
as integrals of motion for integrable system associated with
matrix $T(\lambda)$. For instance, representation (\ref{Lax-XXX})
generates one linear and one quadratic integrals of motion in
variables $s_i,t_i$ and the corresponding integrable system  is
equivalent to a special case of Poincar\'{e} system \cite{poi01}.

According to \cite{soktsig2}, we can construct  commutative
subalgebra generated by {\it quad\-ra\-tic} polynomials on
coefficients of $T_{ij}(\lambda)$. Let us introduce the matrix
\bq
\widetilde{T}(\lambda)=\mathcal K_d(\lambda)\,T(\lambda)\,,
\label{newlax}
\eq
where $T(\lambda)$ is given by (\ref{Lax-XXX}) and
\bq\label{dK}
\mathcal  K_d(\lambda)=\left(\begin{array}{cc}
\lambda+ \mathcal A_0 & a_1\lambda+a_0 \\
 b_1\lambda+b_0 & 0
\end{array}\right).
\eq
Here $a_k,b_k$ are arbitrary numerical parameters and $\mathcal
A_0$ depends on dynamical variables
\[
\mathcal
A_0=a_1(is_2+it_2-t_1-s_1)-b_1(s_1+t_1+is_2+it_2)-s_3-t_3.
\]
We can say that dynamical  matrix $\mathcal K_d$ (\ref{dK})
describes some {\it dynamical} boundary conditions.

The trace of $\widetilde{T}(\lambda)$
\[
\widetilde{\tau}(\lambda)=\mbox{tr}\,\widetilde{T}(\lambda)=\lambda^3+I_1\lambda+I_2
\]
give rise the commutative subalgebra of the Sklyanin brackets
(\ref{rrpoi}). Quadratic and cubic in variables $s_i,t_i$
polynomials $I_{1,2}$ are  integrals of motion for some integrable
system on $so(4)$.

To compare this system with the known examples of integrable
systems on $so(4)$ we introduce two vectors $\mathbf
J=(J_1,J_2,J_3)$ and $\mathbf y=(y_1,y_2,y_3)$ with entries
\bq \label{yJ}
y_i=\varkappa\,(s_i-t_i),\qquad J_i=s_i+t_i,
\eq
which satisfy to the following Lie-Poisson brackets
\begin{equation} \label{bundle}
\bigl\{ J_i\,,J_j\,\bigr\}= \varepsilon_{ijk}\,J_k\,, \qquad
\bigl\{ J_i\,,y_j\,\bigr\}= \varepsilon_{ijk}\,y_k\,,\qquad
\bigl\{ y_i\,,y_j\,\bigr\}=\varkappa^2\varepsilon_{ijk} J_k\,.
\end{equation}
Because physical quantities $y_k,\, J_k$ should be real, parameter
$\varkappa^2$ must be real too and algebra (\ref{bundle}) is
reduced to its two real forms $so(4,\mathbb R)$ or $so(3,1,\mathbb
R)$ for positive and negative $\varkappa^2$ respectfully. The
corresponding Casimir elements are equal to
\bq\label{Caz}
 C_\varkappa=\varkappa^2|\mathbf J|^2+|\mathbf y|^2,\qquad
C=(\mathbf y,\mathbf J).
\eq

Let $(\mathbf y, \mathbf J)$  and $\mathbf y\times \mathbf J$
stand for the inner vector product and for the vector cross
product respectively. In variables (\ref{yJ}) the Hamilton
function is equal to $I_1$ up to constants
\ben
\label{H1} H&=&2I_1+\dfrac{C_\varkappa}{2\varkappa^2}+2\delta_1\delta_2=\\
\nn\\
&=&|\mathbf J|^2+(\mathbf a,\mathbf J)(\mathbf b,\mathbf J)
+\varkappa^{-1}(\mathbf b,\mathbf y\times \mathbf J)-\Bigl(\mathbf
b,\,(\delta_1-\delta_2)\varkappa^{-1}\mathbf
y+(\delta_1+\delta_2)\mathbf J\Bigr)+2(\mathbf c,\mathbf J),\nn
\en
 where
numerical vectors are
\[\mathbf a=\Bigl(0,0,2i\Bigr),\qquad
\mathbf b=\Bigl(i(a_1+b_1),\,a_1-b_1,\,i\Bigr),\qquad \mathbf
c=\Bigl( a_0+b_0,-i(a_0-b_0),\,0 \Bigr)\,.
\]
 Additional integrals of motion $K$ looks like
\ben
K&=&-4iI_2=(\mathbf b,\mathbf J)\Bigl[2|\mathbf
J|^2+\varkappa^{-1}\Bigl(\mathbf a,\mathbf y\times \mathbf
J-(\delta_1-\delta_2)\mathbf y+\varkappa(\delta_1+\delta_2)\mathbf
J\Bigr)-
\varkappa^{-2}C_\varkappa-4\delta_1\delta_2\Bigr]\nn\\
\nn\\
&-&2\varkappa^{-1}(\delta_1-\delta_2)(\mathbf c,\mathbf
y)+2\varkappa^{-1}(\mathbf c,\mathbf y\times \mathbf
J)+2(\delta_1+\delta_2)(\mathbf c,\mathbf J)\label{K1}
\en
Integrals of motion $H$ and $K$ are defined up to canonical
transformations.

Suppose that Hamiltonian function has to depend on the third
component of the vector $\mathbf y\times \mathbf J$ only. The
Hamiltonian (\ref{H1}) has such form after rotation $\mathbf y\to
U\mathbf y$ and $\mathbf J\to U\mathbf J$ on the following Euler
angles
\[
\phi=\dfrac{\pi}{2}-i\dfrac{\ln{a_1}-\ln{b_1}}{2}, \qquad\psi=0,
\qquad\theta=\dfrac{\pi}{2}-i\ln(-i+2\sqrt{a_1b_1})-\dfrac{i}{2}\ln(4a_1b_1+1).
\]
This rotation acts on the numerical vectors ${\mathbf a},{\mathbf
b}$ and $\mathbf c$ in the following way
\[\begin{array}{c}
  \widetilde{\mathbf a}=U^{-1}\mathbf a=\Bigl(0,
2\sqrt{1-c^2},2ic\Bigr),\qquad\widetilde{\mathbf b}=U^{-1}\mathbf
b=\Bigl(0,0,-ic^{-1}), \\
   \\
  \widetilde{\mathbf c}=U^{-1}\mathbf c=\dfrac{1}{2c}\Bigl(\alpha, \beta,
-\dfrac{i\sqrt{1-c^2}}{c}\beta\Bigr), \\
\end{array}
\]
where \[ c=\dfrac{1}{\sqrt{4a_1b_1+1}},\qquad
\alpha=\dfrac{2i(a_1b_0-a_0b_1)}{\sqrt{a_1b_1(4a_1b_1+1)}},\qquad
\beta=\dfrac{-2(a_1b_0+a_0b_1)}{\sqrt{a_1b_1(4a_1b_1+1)}}.
\]
Substituting $\widetilde{\mathbf a},\widetilde{\mathbf b}$ and
$\widetilde{\mathbf c}$ instead vectors ${\mathbf a},{\mathbf b}$
and $\mathbf c$ into (\ref{H1}) and (\ref{K1}) one gets integrals
of motion after rotation. The Hamilton function $H$ (\ref{H1})
after rotation and renormalization
\bq \label{H2}
\widehat{H}=\dfrac{H}{\sqrt{4a_1b_1+1}}=
c(J_1^2+J_2^2-J_3^2)-2\sqrt{1-c^2}J_2J_3+\dfrac{1}{i\varkappa}(y_2J_1-y_1J_2)
+\alpha J_1+ \beta J_2 +\gamma J_3+\delta y_3.
 \eq
depends on five essential parameters $c,\alpha,\beta$ and
\[
 \gamma=-i(\delta_1+\delta_2)+4\dfrac{a_1b_0+a_0b_1}{4a_1b_1+1},\qquad
\delta=\dfrac{\delta_2-\delta_1}{i\varkappa}\,.
\]
It  is a real function on $so(3,1,\mathbb R)$  with negative
$\varkappa^2$ only.

According to \cite{skl95,soktsig2}, the separated coordinates
$q_{1,2}$ for (\ref{H2}) are zeros of the polynomial
\bq
T_{11}(\lambda)=(\lambda-q_1)(\lambda-q_2)=0, \label{serGgch1}
\eq
whereas the conjugated momenta are equal to
\bq
p_k=-i\ln {T}_{21}(q_k)-\ln(a_1\,q_k+a_0). \label{serGgch3}
\eq
We can prove that $q_k,p_k$ are Darboux variables using
(\ref{serGgch1}-\ref{serGgch3}) and  brackets (\ref{rrpoi}).

By definition the generating function of integrals of motion is
\ben
\widetilde{\tau}(\lambda)&=&\mbox{trace}\,\widetilde{T}(\lambda)=
\lambda^3+I_1\lambda-I_2= \nn\\
\nn\\
&=&(\lambda+\mathcal A_0)\,T_{11}(\lambda)+(a_1\lambda+a_0)\,
T_{21}(\lambda)+(b_1\lambda+b_0)\,T_{12}(\lambda) \nn
\en
Substituting $\lambda=q_k$ into this equation one gets  two
separated equations
\[
q_k^3+I_1\,q_k+I_2=\exp(ip_k)+\mbox{det}\,\widetilde{T}(q_k)\,\exp(-ip_k),\qquad
k=1,2.
\]
Here we took into account that $T_{11}(q_k)=0$ and
$T_{12}(q_k)=\mbox{det}\,T(q_k)\,T_{21}^{-1}(q_k)$.

\section{Quartic integrals of motion}
According to \cite{skl88}, we can construct another commutative
subalgebra generated by {\it quad\-ra\-tic} polynomials on
coefficients of $T_{ij}(\lambda)$, which are integrals of motion
for another integrable system associated with the same matrix
$T(\lambda)$ (\ref{Lax-XXX}).

Let ${\cal K}_\pm(\lambda)$ be generic numerical solutions of the
reflection equation \cite{wega94}
\bq \label{Kpm} {\cal K}_+=\left(\begin{array}{cc}
b_3 \lambda+\alpha  & (b_1 +i b_2)\lambda\\
 (b_1 - ib_2) \lambda& -b_3\lambda+\alpha
\end{array}\right)\,,
\qquad{\cal K}_-=\left(\begin{array}{cc}
a_3\lambda+ \beta & (a_1 + i a_2) \lambda\\
 (a_1 -ia_2) \lambda&-a_3\lambda+ \beta
\end{array}\right).
\eq
The Lax matrix for the two-site $XXX$ Heisenberg magnet with
boundaries has the form
\bq \label{refLax}
L(\lambda)={\cal K}_+(\lambda)\,T(\lambda)\,{{\cal
K}}_-(\lambda)\,\left(\begin{array}{cc}
 0 & 1 \\
 -1 & 0
\end{array}\right)T^t(-\lambda)\left(\begin{array}{cc}
 0 & 1 \\
 -1 & 0
\end{array}\right)\,.\label{Laxref}
\eq
Trace of $L(\lambda)$
\bq
\tau(\lambda)=\mbox{\rm tr}\,L(\lambda)=-2(\mathbf a,\mathbf
b)\lambda^6-I_1\lambda^4-I_2\lambda^2-I_3\,.
\eq
give rise to the commutative subalgebra of the Sklyanin brackets
(\ref{rrpoi}). Integrals of motion $I_1,I_2$ and $I_3$ are second,
fourth and sixth order polynomials in variables $s_i,t_i$.

In variables (\ref{yJ}) the Hamilton function $H$ (\ref{H_ini}) is
equal to $I_1$ up to constants
\ben\label{H_ini}
  H&=&I_1+(\mathbf a,\mathbf
  b)\Bigl(\varkappa^{-2}\,C_\varkappa-2(\delta_1^2+\delta_2^2)\Bigr)
   -2 \alpha  \beta=\nn\\
   \nn\\
&=&(\mathbf J,A\mathbf J)+{\varkappa}^{-1}(\mathbf a\times \mathbf
b,\mathbf
y\times\mathbf J) +2(\alpha  \mathbf a+\beta \mathbf b,J)\\
\nn\\
&-&{\varkappa}^{-1}(\delta_1-\delta_2)(\mathbf a\times \mathbf
b,\mathbf y)-(\delta_1+\delta_2)(\mathbf a\times \mathbf b,\mathbf
J),\nn
\en
where $\mathbf a=(a_1, a_2, a_3)$, $\mathbf b=(b_1, b_2, b_3)$ are
numerical vectors and
\bq \label{A_ini}
 A=\mathbf a\otimes\mathbf b+\mathbf b\otimes\mathbf
a,\qquad\qquad A_{ij}=a_ib_j+a_jb_i.
\eq
Additional integral of motion
\bq\label{I_2}
  K=2\varkappa^2I_2-(\mathbf a,\mathbf
  b)\left(C^2+\dfrac{C_\varkappa^2}{4\varkappa^2}\right)+2\alpha\beta C_\varkappa
\eq
is a third order polynomial in momenta $\mathbf J$. For brevity we
present $K$ at $\delta_1=\delta_2=0$ only
\ben\label{K_ini}
K&=&\Bigl(\varkappa^2|\mathbf J|^2-|\mathbf
y|^2\Bigr)\Bigl[\varkappa^{-1}(\mathbf a\times \mathbf b,\mathbf
y\times\mathbf J)+2(\alpha\mathbf a+\beta\mathbf b,J)\Bigr]\\
\nn\\
&+&|\mathbf J|^2\Bigl[8\varkappa^2\alpha\beta-2(\mathbf a,\mathbf
b)|\mathbf y|^2+4\varkappa(\alpha\mathbf a-\beta\mathbf b,\mathbf
y)\Bigr] +(\mathbf y\times\mathbf J,A\mathbf y\times\mathbf
J)\nn\\
\nn\\
&-&4\varkappa(\mathbf y,\mathbf J)(\alpha\mathbf a-\beta\mathbf
b,\mathbf J).\nn
\en
The third coefficient $I_3$ is a constant
\bq\label{I_3}
I_3=\left(\dfrac{C_\varkappa}{4\varkappa^2}+\dfrac{C}{2\varkappa}+\delta_1^2\right)
\left(\dfrac{C_\varkappa}{4\varkappa^2}-\dfrac{C}{2\varkappa}+\delta_2^2\right).
\eq
Integrals of motion (\ref{H_ini}) and (\ref{K_ini}) depend on ten
numerical parameters $a_i,b_i,\alpha,\beta,\delta_1,\delta_2$ and
they are defined up to canonical transformations.

Suppose that Hamiltonian function has to depend on the third
component of the vector $\mathbf y\times \mathbf J$ only. The
Hamiltonian (\ref{H_ini}) has such form after rotation $\mathbf
J\to U\mathbf J$ and $\mathbf y\to U\mathbf y$ with  orthogonal
matrix $U$ defined by
\bq\label{rel}
\widetilde{\mathbf a}=U^{-1}\mathbf
a=\left(\sqrt{\dfrac{e_1}{2}\,},
i\sqrt{\dfrac{e_2}{2}\,},0\right)\,,\qquad \widetilde{\mathbf
b}=U^{-1}\mathbf b=\left(\sqrt{\dfrac{e_1}{2}\,},
-i\sqrt{\dfrac{e_2}{2}\,},0\right),
\eq
where $e_i$ are eigenvalues of the  matrix $A$ (\ref{A_ini}).
Substituting $\widetilde{\mathbf a},\widetilde{\mathbf b}$ instead
vectors ${\mathbf a},{\mathbf b}$ into (\ref{H_ini}) and
(\ref{K_ini}) one gets integrals of motion after rotation.

If  $\mathbf a$ and $\mathbf b$ are linearly dependent vectors,
then $\mathbf a\times \mathbf b=0$ and  matrix $A$ (\ref{A_ini})
has only one non-zero value $e_1\simeq|\mathbf a|^2$. In this case
the Hamilton function (\ref{H_ini})
\[
\widetilde{H}=J_1^2+cJ_1,\qquad c\in{\mathbb R}
\]
determines degenerate or superintegrable system with a
noncommutative family of additional integrals of motion. For
instance, there is the following quadratic integral
\[
\widetilde{K}=c_1J_1^2+J_2^2+J_3^2+c_2y_1^2+c_3(y_2^2+y_3^2)+c_4y_1J_1+c_5(y_2J_3-y_3J_2).
\]
It is special case of the Poincar\'{e} model \cite{poi01}. For
more details see \cite{ts02}.

If $\mathbf a\times \mathbf b\neq0$, the  matrix $A$ (\ref{A_ini})
has two non-zero eigenvalues
\[
e_{1,2}=(\mathbf a,\mathbf b)\pm\,|\mathbf a||\mathbf b|,\qquad
e_3=0.
\]
In this case after  rotation and renormalization the Hamiltonian
$H$ (\ref{H_ini}) is equal to
\ben
\widetilde{H}&=&\dfrac{H}{i\sqrt{e_1e_2\,}}=
cJ_1^2-c^{-1}J_2^2+\varkappa^{-1}\Bigl(y_1J_2-y_2J_1\Bigr)\nn\\
\label{H_fin}\\
&+&\widetilde{\alpha}J_1+\widetilde{\beta}J_2-(\delta_1+\delta_2)J_3-\varkappa^{-1}(\delta_1-\delta_2)y_3\nn
\en
where
\[c=-i\sqrt{e_1e_2^{-1}\,},\qquad
 \widetilde{\alpha}=c_1=-i\sqrt{2e_2^{-1}\,}(\alpha+\beta),\qquad
 \widetilde{\beta}=\sqrt{2e_1^{-1}\,}(\alpha-\beta).
\]
The Hamilton function $\widetilde{H}$ (\ref{H_fin}) depends on
five  parameters  instead of ten parameters in the initial
Hamiltonian (\ref{H_ini}). It allows  us to impose constraints on
the vectors $\mathbf a$ and $\mathbf b$ and to use triangular
boundary matrices $\mathcal K_\pm$ \cite{skl88} instead of the
general ones (\ref{Kpm}). The similar facts hold true  for the
$BC_n$ Toda lattices  \cite{soktsig2} and for the
Kowalevski-Goryachev-Chaplygin gyrostat \cite{ts02b}, which are
also related with the reflection equations.

Thus we can  consider low triangular solution of the reflection
equations $\mathcal K_{+}$ (\ref{Kpm})
\bq \label{rel}
b_1-ib_2=0
\eq
without loss of generality. According to the Sklyanin method
\cite{skl95} the separated variables may be defined by entries of
the following intermediate matrix
\bq {\cal T}(\lambda)=
\,T(\lambda)\,{\mathcal K}_-(\lambda)\,\left(\begin{array}{cc}
  0 & 1 \\
  -1 & 0
\end{array}\right)T^t(-\lambda)\left(\begin{array}{cc}
  0 & 1 \\
  -1 & 0
\end{array}\right),\label{calT}
\eq
which satisfies to the reflection equation \cite{skl88}
\bq
\{\,\on{\cal T}{1}(\lambda),\,\on{\cal T}{2}(\nu)\}=
\left[r(\lambda-\nu),\, \on{\cal T}{1}(\lambda)\on{\cal
T}{2}(\nu)\,\right] +\on{\cal T}{1}(\lambda)r(\lambda+\nu)\on{\cal
T}{2}(\nu) -\on{\cal T}{2}(\nu)r(\lambda+\nu)\on{\cal
T}{1}(\lambda) \,. \label{repoi}
\eq
The separated coordinates $q_{1,2}$ are non-trivial zeros of the
polynomial
\bq
{\cal
T}_{12}(\lambda)=\lambda(\lambda^2-q_1^2)(\lambda^2-q_2^2)=0,
\label{serKgch1}
\eq
whereas the conjugated momenta are equal to
\bq
p_k=-i\ln {\cal T}_{11}(q_k)-\ln(b_3q_k+\alpha). \label{serKgch3}
\eq
We can prove that $q_k,p_k$ are Darboux variables using
(\ref{serKgch1}-\ref{serKgch3}) and the reflection equation
(\ref{repoi}).

Such us $\mathcal K_+$ is a low triangular matrix the generating
function of integrals is
\ben
\tau(\lambda)&=&\mbox{trace}\,L(\lambda)=-2(\mathbf a,\mathbf
b)\lambda^6-I_1\lambda^4-I_2\lambda^2-I_3
= \nn\\
\nn\\
&=&\mbox{trace}\,\mathcal K_+\mathcal
T(\lambda)=(b_3\lambda+\alpha)\mathcal
T_{11}(\lambda)+(b_1+ib_2)\lambda\mathcal
T_{12}(\lambda)-(b_3\lambda-\alpha)\mathcal T_{22}(\lambda) \nn
\en
Substituting $\lambda=q_k$ into this equation one gets two
separated equations
\[
2(\mathbf a,\mathbf
b)\,q_k^6+I_1\,q_k^4+I_2\,q_k^2+I_3=\exp(ip_k)+\mbox{det}\,L(q_k)\,\exp(-ip_k),\qquad
k=1,2.
\]
Here we took into account that $\mathcal T_{12}(q_k)=0$ and
$\mathcal T_{22}(q_k)=\mbox{det}\,\mathcal T(q_k)\,\mathcal
T_{11}^{-1}(q_k)$.

According to \cite{ts02b} we can introduce another separated
variables related with the  proposed separated variables by
canonical transformation and by flip of parameters. Existence of
the different separated variables is associated with the
invariance of the Sklyanin brackets with respect to a matrix
transposition $T\to T^t$.

\section{Conclusion}
The Hamiltonians (\ref{H2}) and (\ref{H_ini}) belongs to the
following class of the Hamiltonians
\bq \label{H_dif}
H=(\mathbf J,A\mathbf J)+(\mathbf a, \mathbf y\times\mathbf J) +
(\mathbf b, \mathbf J)+(\mathbf c, \mathbf y),\nn
\eq
possessing additional integrals of third and fourth degree in
momenta. V.V. Sokolov kindly informed us that there exist three
different such integrable cases only \cite{sok03}.

The research was partially supported by RFBR grant 02-01-00888.

\end{document}